\documentclass[aps,prl,twocolumn,float,showpacs]{revtex4}
\usepackage{amsmath,bm,epsfig}

\let\*\cdot
\def\<{\left\langle} \def\>{\right\rangle} \def\({\left(} \def\){\right)}
 \let\~\widetilde \let\^\widehat 

\def\be{\begin{equation}}\def\ee{\end{equation}}
\def\bea{\begin{eqnarray}}\def\eea{\end{eqnarray}}
\def\bse{\begin{subequations}}\def\ese{\end{subequations}}
\newcommand{\BE}[1]{\begin{equation}\label{#1}}
\newcommand{\BEA}[1]{\begin{eqnarray}\label{#1}}
\newcommand{\BSE}[1]{\begin{subequations}\label{#1}}

\usepackage{pstricks}
\usepackage{pst-node}
\usepackage[ansinew]{inputenc}
\usepackage{amssymb,amsmath}

\def\BSE{\begin{subequations}}\def\ESE{\end{subequations}}
\let \= \equiv

\def\a{\alpha}

\def\be{\begin{equation}}       \def\ba{\begin{array}}

\def\ee{\end{equation}}         \def\ea{\end{array}}

\def\bea {\begin{eqnarray}}      \def\eea {\end{eqnarray}}

\def\bean{\begin{eqnarray*}}    \def\eean{\end{eqnarray*}}

\def\<{\langle} \def\({\left(}  \def\>{\rangle} \def\){\right)}

\newtheorem{exi}{Example}

\begin{document}

\title{Dynamics of nonlinear resonances in Hamiltonian systems}
\author{Miguel D. Bustamante$^{\dag}$  and Elena Kartashova}
 \email{mig_busta@yahoo.com, lena@risc.uni-linz.ac.at}
\affiliation{$^\dag$ Mathematics Institute, University of Warwick, Coventry CV4 7AL, UK \\
$^*$ RISC, J.Kepler University, Linz 4040, Austria}

\begin{abstract}

It is well known that the dynamics of a Hamiltonian system depends
crucially on whether or not it possesses nonlinear resonances. In
the generic case, the set of nonlinear resonances consists of
independent clusters of resonantly interacting modes, described by a
few low-dimensional dynamical systems. We show that 1) most
frequently met clusters are described by integrable dynamical
systems, and 2) construction of clusters can be used as the base for
the Clipping method, substantially more effective for these systems
than the Galerkin method. The results can be used directly for
system with cubic Hamiltonian.
\end{abstract}



\maketitle

 \noindent \textbf{1. Introduction.} A notion of resonance runs through all our life.  Without resonance
we wouldn't have radio, television, music, etc. The general
properties of linear resonances are quite well-known; their
nonlinear counterpart is substantially less studied though interest
in understanding nonlinear resonances is enormous. Famous
experiments of Tesla show how disastrous resonances can be: he
studied experimentally vibrations of an iron column which ran
downward into the foundation of the building, and caused sort of a
small earthquake in Manhattan, with smashed windows and swayed
buildings \cite{Tesla}. Another example is Tacoma Narrows Bridge
which tore itself apart and collapsed (in 1940) under a wind of only
42 mph, though designed for winds of 120 mph.

Nonlinear resonances are ubiquitous in physics. Euler equations, regarded with various boundary conditions and specific values of some
parameters, describe an enormous number of nonlinear dispersive wave systems (capillary waves, surface water waves, atmospheric planetary waves,
drift waves in plasma, etc.) all possessing nonlinear resonances \cite{zakh92}. Nonlinear resonances appear in a great amount of typical
mechanical systems such as an infinite straight bar, a circular ring, and a flat plate \cite{Mech}. The so-called ``nonlinear resonance jump",
important for the analysis of a turbine governor positioning system of hydroelectric power plants, can cause severe damage to the mechanical,
hydraulic and electrical systems \cite{Turbine}. It was recently established that nonlinear resonance is the dominant mechanism behind outer
ionization and energy absorption in near infrared laser-driven rare-gas or metal clusters \cite{Laser}. The characteristic resonant frequencies
observed in accretion disks allow astronomers to determine whether the object is a black hole, a neutron star, or a quark star \cite{Astr}.
Thermally induced variations of the helium dielectric permittivity in superconductors are due to microwave nonlinear resonances \cite{Helium}.
Temporal processing in the central auditory nervous system analyzes sounds using networks of nonlinear neural resonators \cite{Large}. The
non-linear resonant response of biological tissue to the action of an electromagnetic field is used to investigate cases of suspected disease or
cancer \cite{cancer}.

The very special role of resonant solutions of nonlinear ordinary differential equations (ODEs) has been first investigated by Poincar\'e at the
end of the 19th century. Poincar\'e proved that if a nonlinear ODE has no resonance solutions, then it can be linearized by an invertible change
of variables (for details see \cite{Arn3} and refs. therein). This simplifies both analytical and numerical investigations of the original
nonlinear equation, allows for the introduction of corresponding normal forms of ODEs, etc. In the middle of the 20th century, Poincar\'e's
approach has been generalized to the case of nonlinear partial differential equations (PDEs) yielding what is nowadays known as
 KAM theory (\cite{Arn1}--\cite{Mos} and others). This theory allows us to transform a nonlinear dispersive PDE into a Hamiltonian equation
 of motion in Fourier space \cite{lvov},
\bea i\,\dot a_{\bf k} &=& \partial {\cal H}/\partial a_{\bf k}^*,
\label{HamiltonianEquationOfMotion} \eea
 \noindent where $a_{\bf k}$
is  the amplitude of the Fourier mode corresponding to the
wavevector ${\bf k}$ and the Hamiltonian ${\cal H}$ is represented
as an expansion in powers ${\mathcal{H}}_j$ which are proportional
to the product of $j$ amplitudes $a_{\bf k}$. In this Letter we are
going to consider expansions of Hamiltonians up to third order in
wave amplitude, i.e. a cubic Hamiltonian of the form %
\bea {\cal H}_3= \sum_{{\bf k_1}, {\bf k_2},{\bf k_3}} V^1_{23}
a_{1}^* a_2 a_3\delta^1_{23}+\mbox{ complex conj.}, \nonumber
\label{QubicHamiltonian} \eea %
 \noindent  where for brevity we
introduced the notation $ a_j \equiv a_{{\bf k}_j}$ and $ \delta^1_{23} \equiv \delta({\bf k_1} - {\bf k_2} - {\bf k_3}) $ is the Kronecker
symbol. If ${\cal H}_3 \ne 0$, three-wave process is dominant and the main contribution to the nonlinear evolution comes from the
waves satisfying the following resonance conditions: %
 \bea \nonumber
\label{res}
\begin{cases}
\omega ({\bf k}_1) + \omega ({\bf k}_2)- \omega ({\bf k}_{3}) = \Omega,\\
{\bf k}_1 + {\bf k}_2 -{\bf k}_{3} = 0, \label{resn}
\end{cases}
\eea %
 \noindent  where $\omega({ \bf k})$ is a
dispersion relation for the linear wave frequency and $\Omega\ge0$
is called resonance width.

\noindent If $\Omega>0$, the equation of motion (\ref{HamiltonianEquationOfMotion})
turns into%
 \bea \label{Width} i\dot A_{\bf k} = \omega_{\bf
k} A_{\bf k} + \sum_{{\bf k_1}, {\bf k_2}} V^{\bf k}_{12} A_{1} A_2
\delta^{\bf k}_{12}+ 2 V^{1 \, *}_{{\bf k}2}  A_1 A_{2}^* \delta^1_{{\bf
k}2}.
\label{EquationOfMotion3W} \eea%
 If $\Omega=0$, the equation of motion
(\ref{HamiltonianEquationOfMotion}) turns into%
 \bea \label{NoWidth}
i\dot B_{\bf k} =&  \sum_{{\bf k_1}, {\bf k_2}} \big( V^{\bf k}_{12} B_{1} B_2 \delta^{\bf k}_{12} \delta (\omega_{\bf k} -\omega_1 - \omega_2)
 \nonumber \\
&+ 2 V^{1 \, *}_{{\bf k}2}  B_1 B_{2}^*  \delta^1_{{\bf k}3} \delta (\omega_1 -\omega_{\bf k} - \omega_2)\big). \label{EquationOfMotion3Wd}\eea

 \noindent The co-existence of these two substantially different types of wave
interactions, described by Eqs.(\ref{EquationOfMotion3W}) and
(\ref{EquationOfMotion3Wd}), has been observed in numerical
simulations \cite{z2} and proven analytically in the frame of the
kinematic two-layer model of laminated turbulence \cite{K06-Lam}.
Dynamics of the layer (\ref{Width}) is described by wave kinetic
equations and is well studied \cite{lvov}. Dynamics of the layer
(\ref{NoWidth}) is practically not studied though a lot of
preliminary results is already known. Namely, the layer
(\ref{NoWidth}) is described by a few independent wave clusters
formed by the waves which are in exact nonlinear resonance
\cite{AMS}. The corresponding solutions of (\ref{resn}) can be
computed by the specially developed q-class method, presented in
\cite{K06-3} and implemented in \cite{KK06-07}. The general form of
dynamical systems describing resonant clusters can also be found
algorithmically \cite{KM07}, as well as coefficients of dynamical
systems \cite{K-all08}. Moreover, as it was demonstrated in
\cite{PRL94} (numerically) and in \cite{K07} (analytically), these
clusters ``survive" for small enough but non-zero $\Omega.$ The main
goal of this Letter is to study the dynamics of the most frequently
met resonant clusters.\\

 \noindent \textbf{2. Clusters.}
In this Letter we present some analytical and numerical results for
the three most commonly met dynamical systems corresponding to
non-isomorphic clusters of nonlinear resonances -- \emph{a triad},
\emph{a kite}, and \emph{a butterfly} consisting of 3, 4 and 5
complex variables correspondingly.

\noindent The dynamical system for \emph{a triad} has the form
 \be \label{dyn3waves}  \dot{B}_1=   Z B_2^*B_3,\quad
\dot{B}_2=   Z B_1^* B_3, \quad \dot{B}_3= -  Z B_1 B_2. %
\ee 

 \noindent \emph{A kite} consists  of two  triads  $\ a\ $ and $\ b,\ $ with
wave amplitudes
 $\ B_{ja}, \ B_{jb},\ $ $j=1,2,3$,  connected {\it via} two common
 modes. Analogously to \cite{KL-08}, one can point out 4 types of
 kites according to  the properties of connecting modes. For our considerations,
 this is not important: the general method to study integrability of
 kites will be the same. For the concreteness of presentation, in
 this Letter
\emph{a kite} with
 $\ B_{1a}=B_{1b}\ $ and $\ B_{2a}=B_{2b}\ $ has been chosen:
 \bea \label{PP-PP}
\begin{cases}
\dot{B}_{1a}=  B_{2a}^* (Z_a B_{3a} +   Z_b B_{3b}) \,, \\
\dot{B}_{2a}=  B_{1a}^* (Z_a B_{3a} +  Z_b B_{3b}) \,, \\
\dot{B}_{3a}=  - Z_{a} B_{1a} B_{2a} \,,\    \dot{B}_{3b}=  - Z_{b} B_{1a} B_{2a}\ . \\
\end{cases}
 \eea%

 \noindent \emph{A butterfly} consists  of two  triads  $\ a\ $ and $\ b,\ $
with wave amplitudes
 $\ B_{ja}, \ B_{jb},\ $ $j=1,2,3$,  connected {\it via} one common mode. As it was shown in \cite{KL-08},
 there exist 3 different types of butterflies, according to the choice of the connecting
 mode.
Let us take, for instance, $\ B_{1a}=B_{1b} (\equiv B_1).\ $ The corresponding dynamical system is then as follows:
 \bea \label{PP}
\begin{cases}
\dot{B}_{1}=  Z_a B_{2a}^*B_{3a} +   Z_b B_{2b}^*B_{3b}\,, \\
\dot{B}_{2a}=  Z_a B_{1}^* B_{3a}\,,\quad    \dot{B}_{2b}=  Z_b B_{1}^* B_{3b}\,, \\
\dot{B}_{3a}=  - Z_{a} B_{1} B_{2a} \,,\    \dot{B}_{3b}=  - Z_{b} B_{1} B_{2b}\ . \\
\end{cases}
 \eea%

\noindent \textbf{3. Integrability of resonance clusters.} From here on, general notations and terminology will follow Olver's book \cite{Olv93}.
We use hereafter Einstein convention on repeated indices and $f_{,i} \equiv \partial f/\partial x^i$. Consider a general $N$-dimensional system
of autonomous evolution equations of the form:
 \be \label{m: evol}\frac{d{x}^i}{dt}(t) = \Delta^i(x^j(t)), \quad i=1, \ldots, N.
  \ee
 Any scalar function $f(x^i,t)$ that
 satisfies $\frac{d}{dt}\left(f(x^i(t),t)\right) = \frac{\partial}{\partial t} f + \Delta^i f_{,i}  = 0$
is called {\it a conservation law} in \cite{Olv93}. It is easy to see that this definition gives us two types of conservation laws.
The first
type is the standard notion used in classical physics: the conservation law is of the form $f(x^i)$, i.e. it does not depend
explicitly on time.
The second type looks more like a mathematical trick: it is of the form $f(x^i,t)$, where the time dependence is explicit.
In this Letter we will
be interested in both types of conservation law. We claim that they are both physically important and to show that we present
an illustrative
example from Mechanics. Let us regard the damped harmonic oscillator. The equations of motion in non-dimensional form can
be written as:

\be \label{eq:DHO} \dot{q}=p\,, \quad \dot{p} = -q - \alpha p, \ee

where $\a \geq 0$ is the damping coefficient. If this coefficient is equal to zero, $\a=0$, then the total energy of the system

$$E(q,p) = 1/2 \left(p^2 + q^2\right)$$
 is a conservation law of the first type. If $\a > 0$, then the system does not conserve the energy
anymore but one can still define a conserved quantity which is a generalization of $E$ for the case $\a > 0$.
This new quantity $$F(q,p,t) =
\exp(\a \, t) \left(p^2 + q^2 + \a \,p\, q\right)$$ is a conservation law of the second type. This means that for an arbitrary solution $q(t),
p(t)$ of the system (\ref{eq:DHO}) we have
$$\frac{d}{dt}\left[\exp(\a\, t) \left(p(t)^2 + q(t)^2 + \a \,p(t)
\,q(t)\right)\right] = 0$$ 

These two types of conservation law are very different but
complementary. While the first type, $E(q,p)$ in the case $\a=0$,
defines \emph{where} the motion takes place, the second type,
$F(q,p,t)$ in the case $\a > 0$, defines \emph{how} the motion takes
place. In other words, the first type defines orbits of the
dynamical system (\ref{eq:DHO}) and the second type defines its
motion within the orbit.

To keep in mind the difference between these two types of conservation laws, we will call the first type just a conservation law (CL), and we
will call the second type a \emph{dynamical invariant}.

We say that Sys.(\ref{m: evol}) is {\it integrable} if there are $N$
functionally independent dynamical invariants. Obviously, if
Sys.(\ref{m: evol}) possesses $(N-1)$ functionally independent CLs,
then it is constrained to move along a $1$-dimensional manifold, and
the way it moves is dictated by 1 dynamical invariant. This
dynamical invariant can be obtained from the knowledge of the
$(N-1)$ CLs and the explicit form of the Sys.(\ref{m: evol}), i.e.
Sys.(\ref{m: evol}) is integrable then. It follows from the Theorem
below that in many cases the knowledge of only $(N-2)$ CLs is enough
for the integrability of the Sys.(\ref{m: evol}).\\

{\textbf{Theorem.}}  \emph{Let us assume that the system (\ref{m: evol}) possesses a standard Liouville volume density $$\rho(x^i): \ ( \rho
\Delta^i)_{,i} = 0,$$ and $(N-2)$ functionally independent CLs, $H^1, \ldots, H^{N-2}$. Then a new CL can be constructed, which is functionally
independent of the original ones, and therefore the system is integrable.}\\

 \noindent  The (lengthy) proof follows from the existence of a Poisson bracket
for the original Sys.(\ref{m: evol}) and is an extension of the general approach used in \cite{BH03} for three dimensional first order autonomous
equations. The proof is constructive and allows us to find the explicit form of a new CL for dynamical systems of the form
(\ref{dyn3waves}),(\ref{PP-PP}),(\ref{PP}), etc.\\

In the examples below, we always need to eliminate so-called slave phases, which corresponds to the well-known order reduction in Hamiltonian
systems \cite{MaWe74}. The number $N$ used below corresponds to the effective number of degrees of freedom after this reduction has been
performed.

Integrability of \emph{a triad}, dynamical
system (\ref{dyn3waves}), is a well-known fact (e.g. \cite{book-triad}) and its two conservation laws are%
 \be \nonumber \label{laws3waves}
 I_{23}=|B_2 |^2 + |B_3|^2,  \  I_{13}= |B_1 |^2 + |B_3|^2.
\ee
Sys.(\ref{dyn3waves}) has been used for a preliminary check of our method; in this case $N=4$. The method can thus be applied and we obtain the
following CL:
 \be \nonumber \label{triad-CLs_new} I_T = \operatorname{Im}(B_1 B_{2} B_{3}^*)\,,\ee
 together with the time-dependent dynamical invariant of the form:

 \be \nonumber \label{triad-dyn}
 S_0 = Z\, t
 -\frac{F\left(\arcsin\left(\left({\frac{R_3 - v}{R_3 - R_2}}
 \right)^{1/2}\right),\left({\frac{R_3-R_2}{R_3-R_1}}\right)^{1/2}\right)}{2^{1/2}(R_3-R_1)^{1/2}
 (I_{13}^2 - I_{13}I_{23} + I_{23}^2)^{1/4}} . \nonumber \ee

Here $F$ is the elliptic integral of the first kind, $R_1 < R_2 <
R_3$ are the three real roots of the polynomial  $$x^3 + x^2 = 2/27
-(27 I_T^2 - $$
$$ (I_{13}+I_{23})(I_{13}-2
I_{23})(I_{23}-2 I_{13}))/27 (I_{13}^2 - I_{13}I_{23} +
I_{23}^2)^{3/2}$$
 \noindent  and
$$v = |B_1|^2 - (2I_{13}-I_{23} + (I_{13}^2 - I_{13}I_{23} +
I_{23}^2)^{1/2})/3$$
 \noindent  is always within the interval
$[R_2,R_3] \ni 0.$

\emph{A kite}, dynamical system (\ref{PP-PP}), is also an integrable system. Indeed, after reduction of slave variables the system corresponds to
$N=6$ and has 5 CLs (2 linear, 2 quadratic, 1
cubic):%

 \bea \nonumber \label{kite-CLs}
 \begin{cases}\nonumber
 L_{R}= \operatorname{Re}(Z_b B_{3a} - Z_a B_{3b}), \ L_{I}= \operatorname{Im}(Z_b B_{3a} - Z_a B_{3b})\,,\\
 I_{1ab}= |B_{1a}|^2 + |B_{3a} |^2 + |B_{3b}|^2\,,\\
 I_{2ab}= |B_{2a}|^2 + |B_{3a} |^2 + |B_{3b}|^2\,,\\
 I_K = \operatorname{Im}(B_{1a} B_{2a} (Z_a B_{3a}^*+Z_b
B_{3b}^*))\,,
 \end{cases}
 \eea %

 \noindent with a dynamical invariant that is essentially the same as for a
triad, $ S_0$, after replacing $Z = Z_a+Z_b,  I_T = I_K
(Z_a^2+Z_b^2)/Z^3,$ $I_{13}= I_{1ab} (Z_a^2+Z_b^2)/Z^2 - (L_R^2 +
L_I^2)/Z^2,$ $I_{23}= I_{2ab} (Z_a^2+Z_b^2)/Z^2 - (L_R^2 +
L_I^2)/Z^2.$

The dynamics of  \emph{a butterfly} is governed by Eqs.(\ref{PP}) and its 4 CLs (3 quadratic and 1 cubic) can easily be obtained:

 \bea \label{int-PP}
\begin{cases} I_{23a}=|B_{2a} |^2 + |B_{3a}|^2 \,, \quad
 I_{23b}= |B_{2b} |^2 + |B_{3b}|^2 \,, \\
 I_{ab}= |B_{1}|^2 + |B_{3a} |^2 + |B_{3b}|^2 \,, \\
I_0 = \operatorname{Im}(Z_a B_1 B_{2a} B_{3a}^* + Z_b B_1 B_{2b}
B_{3b}^*) \,,
 \end{cases}
  \eea

 \noindent while a Liouville volume density is $\rho = 1$.
 Notice that all cubic CLs are canonical Hamiltonians for the
respective \emph{triad}, \emph{kite} and \emph{butterfly} systems.
From now on we consider the \emph{butterfly} case when no amplitude is
identically zero; otherwise the system would become integrable.

\begin{figure*}
\begin{center}
\includegraphics[width=5cm,height=5cm,angle=270]{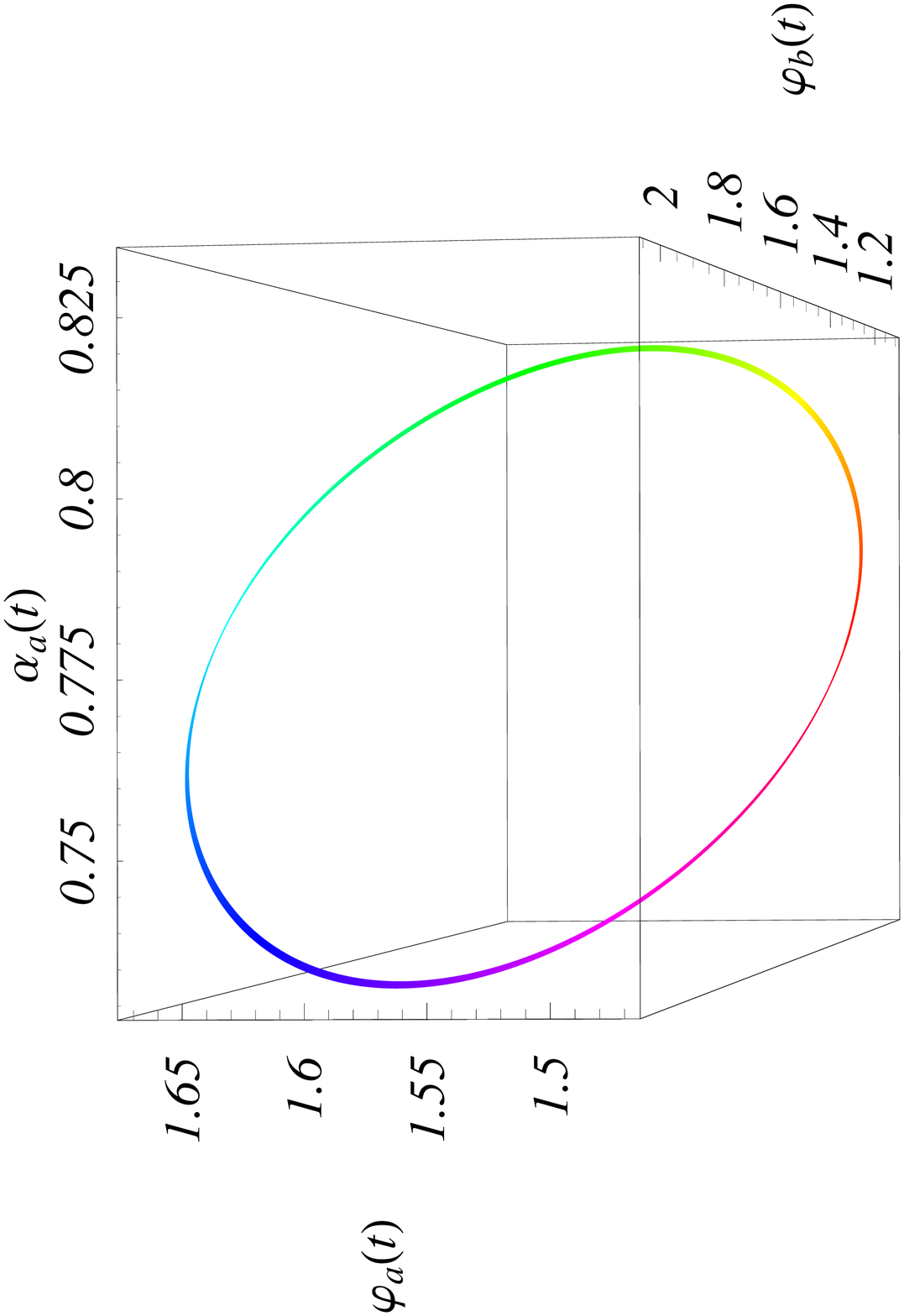}
\includegraphics[width=5cm,height=5cm,angle=270]{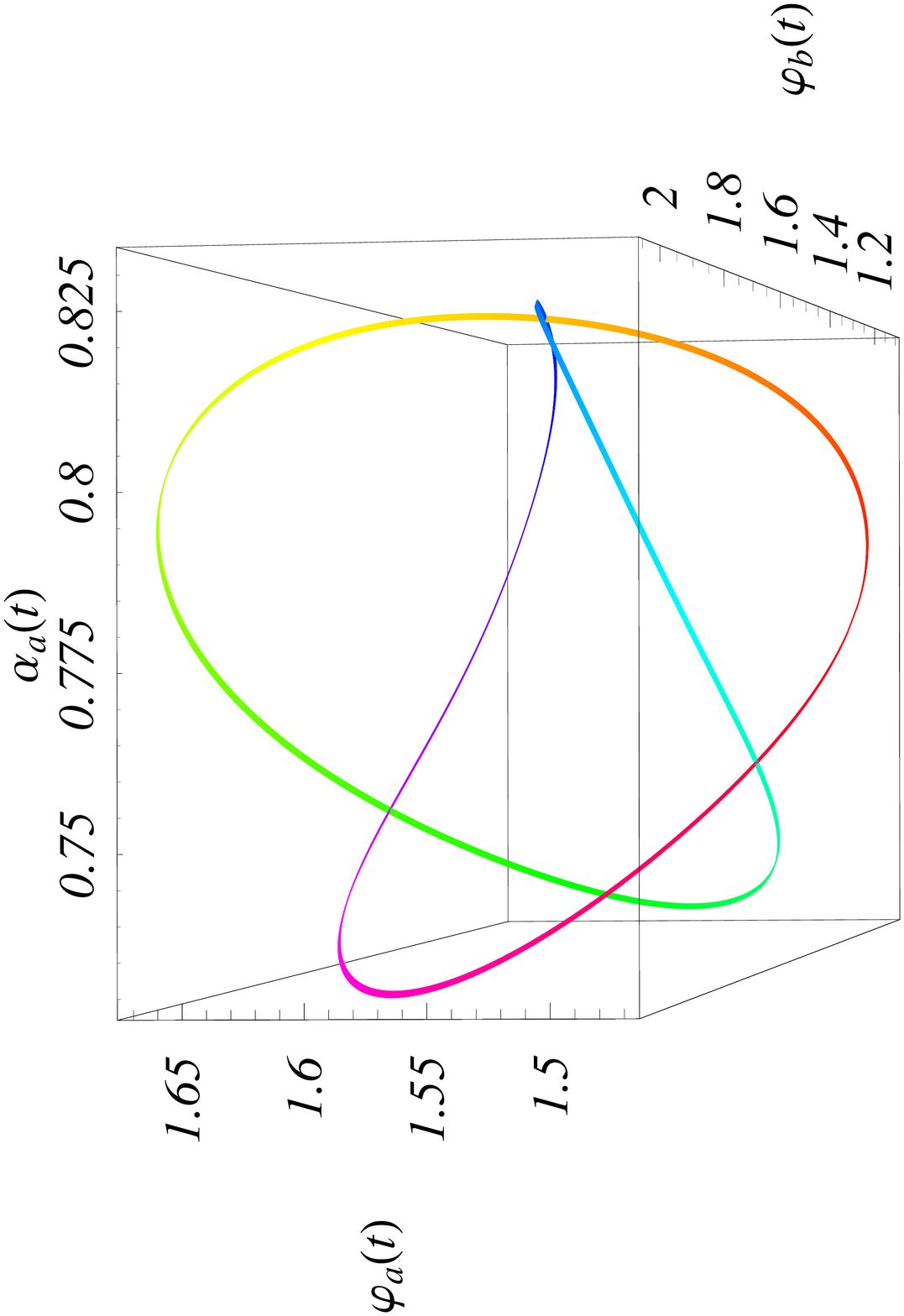}
\includegraphics[width=5cm,height=5cm,angle=270]{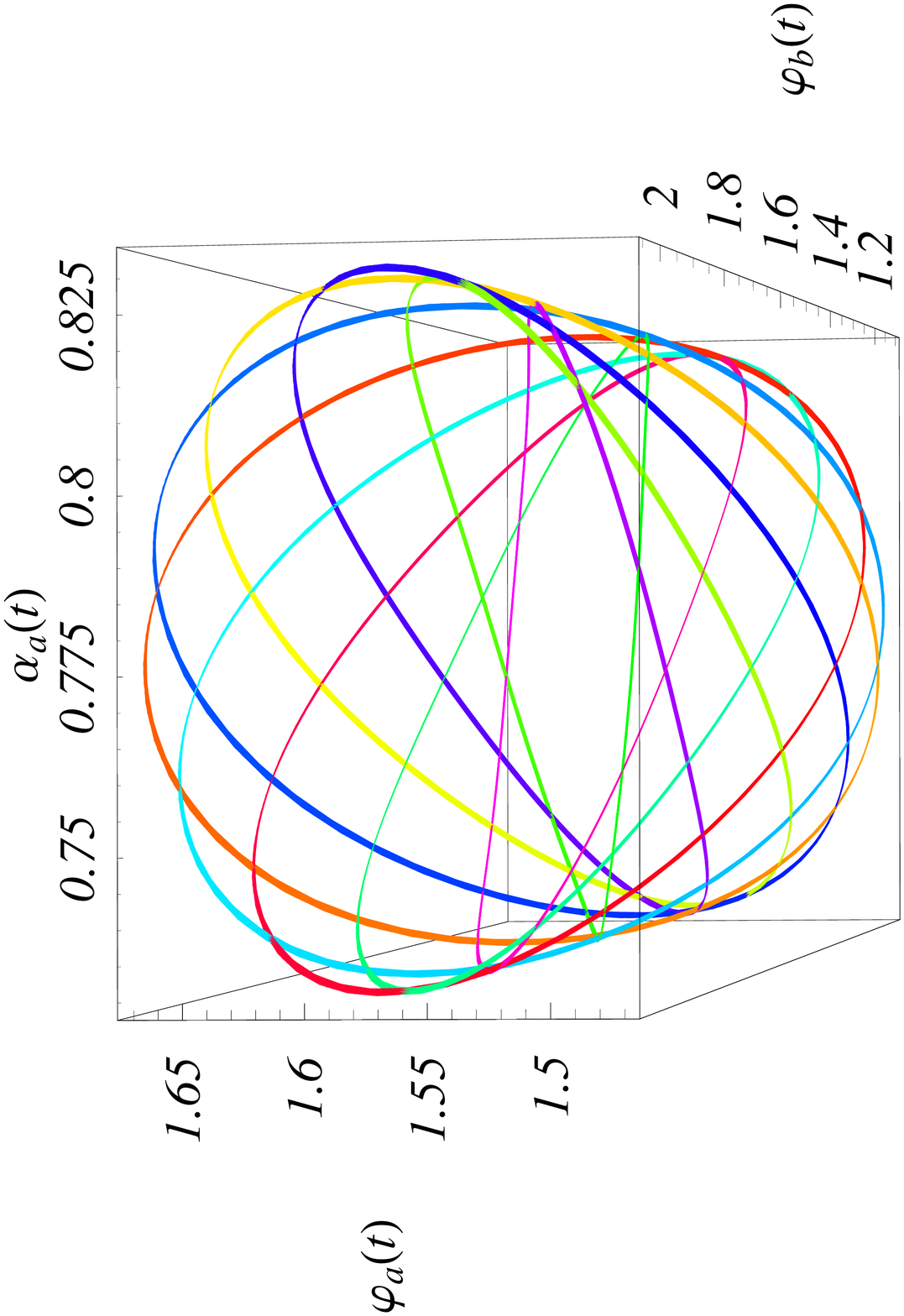}
\end{center}
\caption{\label{f:dynamics} Color online. To facilitate view, color hue of the plot is a linear function of time $t$, varying from 0 to 1 as $t$
runs through one period. {\bf Left panel}: $Z_a = Z_b = 10/100 (integrable case), I_{ab} \approx 2.1608$. {\bf Middle panel}:
 $Z_a = 8/100 , Z_b=12/100, I_{ab} \approx 2.2088$. {\bf Right
panel}: $Z_a = 9/100 , Z_b=11/100, I_{ab} \approx 2.1846$.}
\end{figure*}

 The use of standard  amplitude-phase  representation $B_j= C_j\exp(i
\theta_j)$ of the complex amplitudes $B_j$ in terms of real
amplitudes $C_j$ and phases $\theta_j$ shows immediately  that  only
two phase combinations are important:
 \be \nonumber \label{ab-phases}
 \varphi_{a} = \theta_{1a} + \theta_{2a} -\theta_{3a}, \quad
\varphi_{b} = \theta_{1b} + \theta_{2b} -\theta_{3b}, \ee
$a$- and $b$-triad phases (with  the requirement
$\theta_{1a}=\theta_{1b}$ which correspond to the chosen resonance
condition). This reduces five complex equations (\ref{PP})  to only
four real ones:
  \bea
\label{m: C3a}  \frac{d C_{3a}}{dt}&=& - Z_a C_1  C_{2a} \cos \varphi_a\,, \\
\label{m: C3b} \frac{d C_{3b}}{dt} &=&   - {Z_b} \, C_1  C_{2b} \cos \varphi_b\,,   \\
\label{m: phi_a}\frac{d \varphi_a}{dt}&=&   Z_a C_1 \left(
\frac{C_{2a}}{C_{3a}} - \frac{C_{3a}}{C_{2a}}\right)\sin \varphi_a -
\frac
{I_{0}}{(C_1)^2}\,,\\
\label{m: phi_b}\frac{d \varphi_b}{dt}&=&  {Z_b} C_1 \left(
\frac{C_{2b}}{C_{3b}} - \frac{C_{3b}}{C_{2b}}\right)\sin \varphi_b
- \frac {I_{0}}{(C_1)^2}\,.
 \eea
The cubic CL reads
 \be \label{m: Hamiltonian} I_0 = C_1 \left( Z_a C_{2a} C_{3a} \sin \varphi_a + Z_b
C_{2b} C_{3b} \sin \varphi_b\right)
 \ee
in terms of the amplitudes and phases. This means that the dynamics
of a butterfly cluster is, in the generic case, confined to a
$3$-dimensional manifold. Below we regard  a few particular cases in
which Sys.(\ref{PP}) is integrable.

 \noindent {\bf Example 1: Real amplitudes, $\varphi_a = \varphi_b = 0.$} In this case the
Hamiltonian $I_0$ becomes identically zero while Liouville density
in coordinates $C_{3a}, C_{3b}$ is $\rho(C_{3a}, C_{3b}) = 1/C_1
C_{2a} C_{2b}$.
 So in this case the
equations for the unknown CL $H(C_{3a}, C_{3b})$ are:
 \be \rho \Delta^{C_{3a}} =  -\frac{Z_a}{C_{2b}}  =  \frac{\partial}{\partial C_{3b}} H , \ \
\rho \Delta^{C_{3b}} = -\frac{Z_b}{C_{2a}}  = -
\frac{\partial}{\partial C_{3a}} H, \nonumber
 \ee
and from Eqs.(\ref{int-PP}) we readily obtain
$$H(C_{3a}, C_{3b}) = Z_b \arctan\left(C_{3a}/C_{2a}\right) - Z_a
\arctan\left(C_{3b}/C_{2b}\right),$$ i.e. Sys.(\ref{PP}) is
integrable in this case. Of course, this case is degenerate for $I_0
\equiv 0$ yields no constraint on the remaining independent
variables $C_{3a}, C_{3b}$ satisfying equations (\ref{m: C3a}),
(\ref{m: C3b}).

 \noindent {\bf General change of coordinates}. Going back to Eqs.(\ref{m:
C3a})--(\ref{m: Hamiltonian}), we can jump from this degenerate case to a more generic
case by defining new coordinates which are suggested by
$H(C_{3a}, C_{3b}).$ The new coordinates are to replace the amplitudes $C_{3a}, C_{3b}$:
 \be \nonumber \label{m: new coord} \alpha_a = \arctan\left(C_{3a}/C_{2a}\right),
 \ \
\alpha_b = \arctan\left(C_{3b}/C_{2b}\right)\,.
 \ee
We choose the inverse transformation to be
 \be
 \begin{cases}
 C_{2a} = \sqrt{I_{23a}} \cos(\alpha_a)\,, \quad C_{3a} = \sqrt{I_{23a}}\sin(\alpha_a)\,,\\
 C_{2b} = \sqrt{I_{23b}} \cos(\alpha_b)\,, \quad C_{3b} = \sqrt{I_{23b}}\sin(\alpha_b)\,,
 \end{cases}
 \ee
 so that the domain for the new variables is $0<\alpha_a<\pi/2, \quad 0<\alpha_b <\pi/2$.
 For these new coordinates, the evolution equations simplify
enormously:
  \bea \label{m: alpha}
\begin{cases} \frac{d \alpha_{a}}{dt} = - Z_a C_1 \cos \varphi_a, \ \
 \frac{d \alpha_{b}}{dt} = - Z_b C_1 \cos
\varphi_b\,, \\
\frac{d \varphi_a}{dt}  =   Z_a C_1 \left( \cot \alpha_a - \tan \alpha_a\right)\sin \varphi_a - \frac
{I_{0}}{(C_1)^2} ,\\
\frac{d \varphi_b}{dt}  =  {Z_b} C_1 \left( \cot \alpha_b - \tan \alpha_b\right)\sin \varphi_b  - \frac {I_{0}}{(C_1)^2},
 \end{cases}
 \eea
where the amplitude $C_1>0$ is obtained using eqs.(\ref{int-PP}):
 \be \label{m: amplitude C_1} C_1 = \sqrt{I_{ab} - I_{23a} \sin^2 \alpha_a  - I_{23b} \sin^2 \alpha_b }\,
 \ee
and the cubic CL is now
 \be \label{m:Hamiltonian_final} I_0 = \frac{C_1} 2 \left(Z_a I_{23a} \sin(2\alpha_a) \sin(\varphi_a) + Z_b I_{23b} \sin(2\alpha_b) \sin(\varphi_b)\right).
 \ee
Equations (\ref{m: alpha})--(\ref{m:Hamiltonian_final}) represent the final form of our $3$-dimensional general system.

 \noindent
{\bf Example 2: Complex amplitudes, $I_0=0.$} Here, we just impose the condition $I_0=0$ but the phases are otherwise arbitrary: this case is
therefore not degenerate anymore and we have a $3$-dimensional system which requires the existence of only $1$ CL in order to be integrable: a CL
is $A_a=\sin(2\alpha_a) \sin(\varphi_a),$ which can be deduced from Eqs.(\ref{m: alpha}). Making use of the Theorem, one can find another CL for
this case:  $H_{new}(C_{3a},C_{3b}) = (1+Z_b/Z_a) \arccos\left(\frac{\cos 2 \alpha_a}{\sqrt{1-A_a^2}}\right)-(1+Z_a/Z_b) \arccos\left(\frac{\cos
2 \alpha_b}{\sqrt{1-A_b^2}}\right)$. Obviously $A_a$ and $H_{new}$ are functionally independent, i.e. the case $I_0 = 0$ is integrable.

 \noindent
{\bf Example 3: Complex amplitudes, $Z_a=Z_b.$} In this case a new
CL has the form
 \bea \frac{I_0^2}{Z_a} E &=& C_{2a}^2 C_{3a}^2 + C_{2b}^2 C_{3b}^2 +
 2 C_{2a} C_{3a} C_{2b} C_{3b} \cos(\varphi_a-\varphi_b)\nonumber \\
 &-& C_1^2(C_1^2 - C_{2a}^2+ C_{3a}^2 - C_{2b}^2 + C_{3b}^2)\,,
 \eea
which is functionally independent of the other known constants of motion.
Therefore, according to the Theorem the case $Z_b=Z_a$ is integrable.

The numerical scheme is programmed in \emph{Mathematica} with
\emph{stiffness-switching} method in single precision. For arbitrary
$Z_a, Z_b$  the scheme has been checked by computing $I_0$ from
eq.(\ref{m:Hamiltonian_final}) at all consequent time steps; there
is no noticeable change of $I_0$ up to machine precision.\\

\begin{figure}
\begin{center}
\includegraphics[width=7.5cm,height=3.2cm,angle=0]{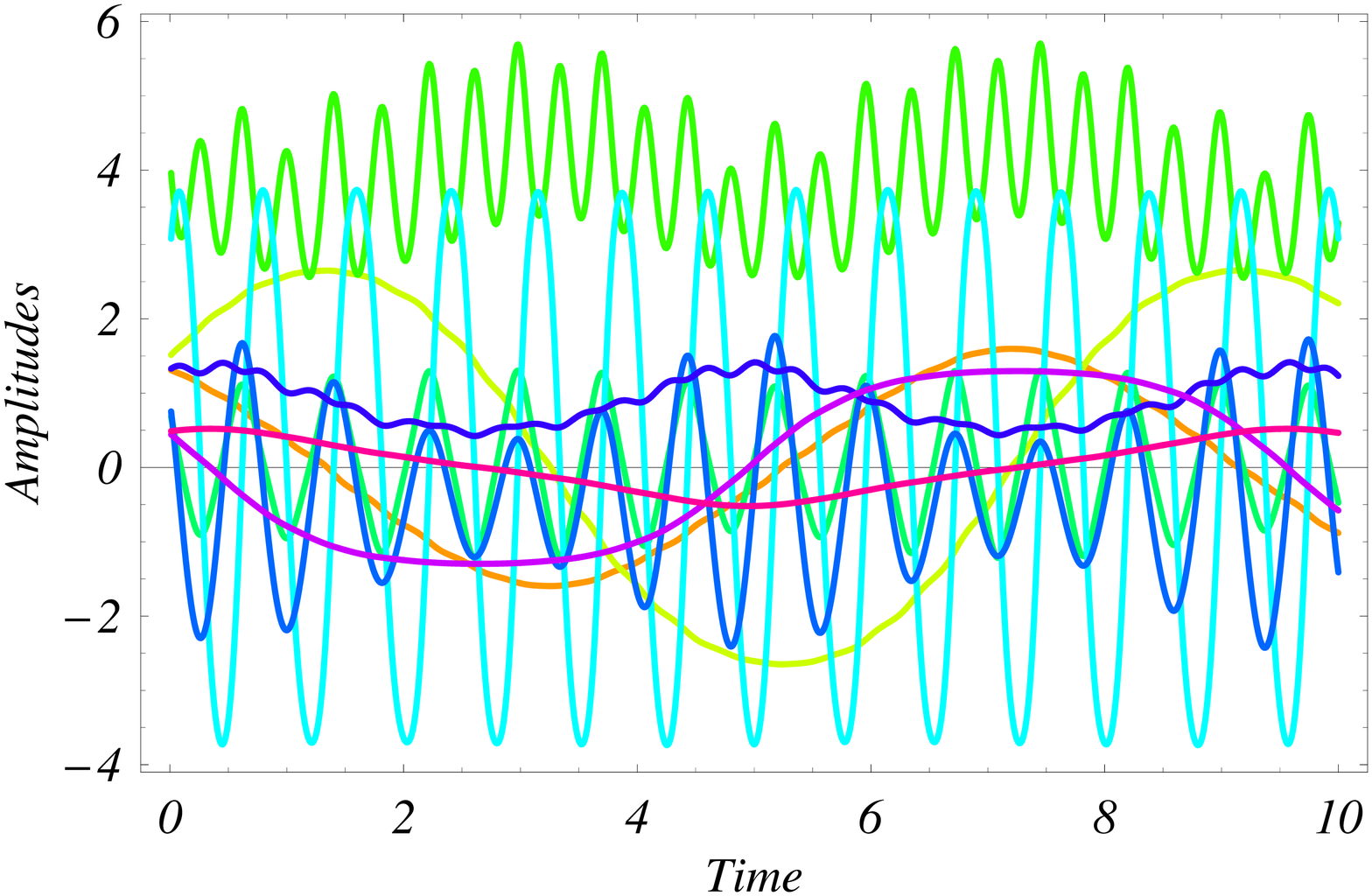}
\end{center}
 \caption{\label{f:dynamics15} Color online. Chain of 4 triads,
 real amplitudes for all 9 modes. Time and amplitudes in non-dimensioned units.}
\end{figure}

\noindent \textbf{4. Numerical simulations.} To investigate the general behavior of a butterfly cluster with $Z_a \neq Z_b$, we integrated
directly Eqs.(\ref{m: alpha}),(\ref{m: amplitude C_1}) with $I_0$ computed from Eq.(\ref{m:Hamiltonian_final}) evaluated at $t=0$ and used to
check numerical scheme afterwards. Some results of the simulations with Sys.(\ref{m: alpha}) are presented in Fig.\ref{f:dynamics}. Initial
conditions $\alpha_a(0) = 78/100, \alpha_b(0) = 60/100, \varphi_a(0) = 147/100, \varphi_b(0) = 127/100$ and values of the constants of motion
$I_0 =11/2000, I_{23a} = 4/100, I_{23b} = 4/100$ are the same for all three parts of Fig.\ref{f:dynamics}. $3$D parametric plots are shown in
space $(\alpha_a, \varphi_a, \varphi_b)$ with color hue depending on time, so the plots are effectively $4$D. The main goal of this series of
numerical simulations was to study changes in the dynamics of a butterfly cluster according to the magnitude of the ratio $\zeta= Z_a/Z_b.$ In
Fig.\ref{f:dynamics}, left panel $\zeta_{l}=1,$ middle panel $\zeta_{m}=2/3$ and right panel $\zeta_{r}=9/11.$ As it was shown above, the case
$\zeta_{l}=1$ is integrable, and one can see the closed trajectory with period $T_{l} \approx 21.7$. Quite unexpectedly, rational $\zeta_{m},
\zeta_{r}$ produce what appear to be periodic motions and closed trajectories with periods $T_{m} \approx 53$ and $T_{r} \approx 215$
correspondingly. A few dozen of simulations made with different rational ratios $\zeta= Z_a/Z_b$ show that periodicity depends - or is even
defined by - the commensurability of the coefficients $Z_a$ and $Z_b$. Fig.\ref{f:dynamics} shows that $\zeta_{m}=2/3$ gives 2 spikes in one
direction and 3 spikes in the perpendicular direction, while $\zeta_{r}=9/11$ gives 9 and 11 spikes correspondingly; and so on.
 This does not mean, of course,
that in the general case $Z_a \neq Z_b$ solutions of Sys.(\ref{PP}) are periodic. But the opposite can only be proven
analytically for in
numerical simulations any choice of the ratio $\zeta$ necessarily turns into a rational number.
Some preliminary series of simulations have been performed with the
chains of triads with connection types as in (\ref{PP}), with
maximum number of triads in a chain being 8 (corresponds to 17
modes). For our numerical simulations, resonant clusters of
spherical Rossby waves were taken, with initial (non-dimensioned)
energies of the order of measured atmospheric data, as in
\cite{KL-06}. The dynamical system for computations was taken in the
original variables $B_j$. The case of real amplitudes is shown in
Fig.\ref{f:dynamics15}: all resonant modes behave (almost)
periodically. Non-dimensioned units for time and amplitudes were
chosen to illustrate clearly the characteristic behavior of the
amplitudes.\\

\noindent \textbf{5. Galerkin method \emph{versus} Clipping method.} Speaking very generally, Galerkin method (GM) allows one to reduce infinite
dimensional systems, described by PDEs, to low dimensional systems of ODEs. Three steps have to be performed: a) choice of ansatz functions
(modes); b) choice of number of modes $N$; c) construction of the resulting reduced system. While in the case of a \emph{linear} PDE, Fourier
modes is the natural choice of the ansatz functions, the number of modes is mostly defined by the available computer facilities and the
construction of the corresponding ODE is obvious, the case of a \emph{nonlinear} PDE is much more involved \cite{BS2005}.

The Clipping method (CM) has been introduced in \cite{TMP94} in order to deal with evolutionary dispersive nonlinear PDEs. The idea of the CM is
very simple: under some physically relevant conditions, it is enough to regard the dynamics of the resonantly interacting modes. These resonant
modes can be found systematically using the q-class method \cite{K06-3}. All other modes behave as linear and can be clipped out. Numerical
evidence of the effectiveness of this approach was presented in \cite{PRL94}, using a pseudo-spectral model of the barotropic vorticity equation.
There, it was observed that the most energetically active modes were the resonant modes (221 modes were excited, among them 39 resonant). These
resonant modes appear in clusters, i.e. a small number of modes, and the energy of each cluster is conserved (see \cite{PRL94}, Fig.1). As for
each of the non-resonant modes, their energies are approximately constant during many periods of energy exchange of the resonant modes
(\cite{PRL94}, Fig.3).

Implementation of the q-class method for various dispersion functions \cite{KK06-07}, together with explicit construction of the dynamical
systems corresponding to the exact resonances \cite{KM07}, allows one to eliminate the arbitrariness of the choices at all three steps a), b) and
c) of the GM. Instead of one big system of ODEs of order $N$, we have a few independent systems of ODEs, of order $\tilde{N} \ll N$, and these
systems are often integrable. Another example can be found in \cite{KM07}: ocean planetary motions, 128 resonant modes among 2500 Fourier
harmonics in the chosen spectral domain, all clusters and their dynamical systems are written out explicitly. In dozens of
studied 3-wave systems, the amount of resonant modes does not exceed $20\%$ of all modes in the spectral domain.\\

\noindent \textbf{6. Conclusions.} Our analysis and general mathematical results~\cite{AMS} on resonant clusters are valid for arbitrary
Hamiltonian ${\cal H}_j, j \geq 3,$ though computation of clusters in the case $j>3$ is more involved \cite{KK06-07}.

Clipping method has at least three advantages compared to Galerkin truncation: i) Numerical schemes in CM can be truncated at a substantially
higher wavenumber than in GM, depending not on the computer facilities but on some physically relevant parameters (say, dissipation range of
wavenumbers). ii) Most of the resulting dynamical systems corresponding to each resonant cluster is analytically integrable. iii) The solutions
obtained from the integrable cases could be used to parameterize the numerical solutions of non-integrable systems found for bigger clusters.
This work is in progress.

Last not least. Even for one specific PDE, it is a highly non-trivial task to prove that Galerkin truncation is a Hamiltonian system and to
construct additional conserved quantity \cite{AKM2003}. Clipping method combined with the constructive procedure based on the Theorem above
(Section 3), allows us to produce physically relevant dynamical systems and to find additional conservation laws
systematically, for a wide class of evolutionary dispersive nonlinear PDEs.\\

 \noindent
{\bf Acknowledgements}. E.K. acknowledges the support of the
Austrian Science Foundation (FWF) under project
 P20164-N18 ``Discrete resonances in nonlinear wave systems". M.B. acknowledge the support of the
 Transnational Access Programme
at RISC-Linz, funded by European Commission Framework 6 Programme
for Integrated Infrastructures Initiatives under the project SCIEnce
(Contract No. 026133).

 \end{document}